# Crystal Structure Studies of Human Dental Apatite as a Function of Age

Th. Leventouri[1*], A. Antonakos[2], A. Kyriacou[1], R. Venturelli[1], E. Liarokapis[2], V. Perdikatsis[3]

[1]Department of Physics and Center for Biological and Materials Physics, Florida Atlantic University, Boca Raton, FL 33431, USA
[2]Department of Physics, National Technical University of Athens, 15780 Athens, Greece
[3]Department of Mineral Resources Engineering, Technical University of Crete, Chania, Greece

**Abstract**

Studies of the average crystal structure properties of human dental apatite as a function of the tooth-age in the range of 5-87 years are reported. The crystallinity of the dental hydroxyapatite decreases with the tooth-age. The $\alpha$-lattice constant that is associated with the carbonate content in carbonate apatite decreases with the tooth-age in a systematic way, whereas the c-lattice constant does not change significantly. Thermogravimetric measurements demonstrate an increase of the carbonate content with the tooth-age. FTIR spectroscopy reveals both, B and A-type carbonate substitutions with the B-type greater than the A-type substitution by a factor up to ~5. An increase of the carbonate content as a function of the tooth-age can be deduced from the ratio of the $\nu_2$ $CO_3$ to the $\nu_1$ $PO_4$ IR modes.

**Introduction**

Hydroxyapatite (HAp) with multiple substitutions at all sites and containing ~4% to 6% carbonate is the primary component of bones (70% wt) and teeth (96% wt) [1, 2]. Several authors have reported on the structure and properties of human dental carbonate HAp [3-5]. They study the enamel part of the tooth with a focus on the crystallographic structure and the carbonate substitution because carbonate affects important properties of the physiological HAp such as reactivity and solubility [6]. Studies on the graded nature and texture of dental enamel by comparison of the microstructures of slices of human adult and baby canine enamel have been reported [7, 8]. Human deciduous and permanent enamel samples were studied by Fourier Transform Infrared (FTIR) spectroscopy to determine quantitatively the B-type (carbonate for phosphate) and A-type (carbonate for hydroxyl) carbonate contents in human enamel [9]. The mineral content, crystallite size and mechanical properties of aging (transparent) human dentin were compared with the ones in normal human dentin in recent studies [10, 11].

We study the average crystal structure properties in bulk human dental apatite as a function of the tooth-age in the range 5-87 years without separating the enamel from the dentin. We have undertaken a research project that requires a large and diverse origin of specimens in order to correlate the average crystal structure properties of aging dental apatite with the parameters that create the structural changes. Understanding the fundamental science of the dental mineral phase as a function of age could be helpful in efforts of remineralisation of human dental apatite [12]. Here we report some preliminary results on systematic trends of average crystal structure parameters and carbonate content in bulk human dental apatite versus the tooth-age by using powder x-ray diffraction (XRD), thermogravimetric analysis (TGA) and FTIR spectroscopy.

**Experimental**

Two local dental offices provided teeth samples for this research with the informed consent of their patients. The teeth-samples were selected in the age range 17-87 years old. One healthy 5 year old deciduous tooth was offered by a family member of the first author. The cleanest pieces of each tooth (free of fillings etc.) were selected under magnification, powdered with an agate mortar and pestle, and passed through a 125 μ sieve. The way of preparation of the samples implies that the average crystal structure properties of the dental apatite are studied. Each specimen was labeled with a capital T followed by a number that represents the tooth-age.

Powder x-ray diffraction measurements were performed using a Siemens D5000 powder diffractometer operating at 45kV and 40mA with Cu-Kα radiation and a diffracted beam monochromator. Data were collected in the 2θ range of $8°$-$90°$ with a step size of $0.02°$ and a counting time of 20 s at each step. The data bank from the International Center for Diffraction Data (ICDD) was used in a search/match program for phase identification. The Rietveld refinement method [13] in the GSAS program [14] was used for crystal structure analysis of the diffraction patterns only for the HAp phase. The crystallographic model used space group $P6_3/m$ with isotropic atomic displacement parameters. First, the scale factor, background, peak profile (pseudovoight function) and lattice parameters were simultaneously refined; then, the atomic positions and isotropic displacement parameters were refined for all the atoms except for the oxygen and hydrogen of the hydroxyl site in the diffraction patterns from the older age teeth, because combination of low occupancy at the channel site, poor crystallization and peak overlapping would create instability of the refinement. For the same reason, the fractions of the Ca1, Ca2 and P atoms were refined in diffraction patterns up to 45 years old teeth.

Thermogravimetric analysis was performed to evaluate the carbonate content in the samples. The loss of weight as a function of temperature from samples of 50-100 mg was recorded using a PerkinElmer thermogravimetric analyzer TGA-6. The heating rate was $10°C$/min in the temperature range 25 $°C$ to 950 $°C$.

FTIR spectroscopy was used to study the carbonate presence in the HAp structure of the specimens. The FTIR spectra were recorded on a Bruker Optic IFS66v/S interferometer equipped with an ATR (Attenuated Total Reflectance) unit. The ATR unit permits the spectra collection without any special sample preparation and it is used for characterization and quantitative estimations in several materials. The range of frequencies was 500 to 4000 cm$^{-1}$ and the spectra were recorded in ambient conditions with a resolution of 2 cm$^{-1}$. In order to obtain a good signal-to-noise ratio, more than a hundred scans were collected and averaged. A KBr beamsplitter was used for the M-IR source.

**Results/Discussion**

Fig. 1 shows the x-ray diffraction pattern collected from the 17 years old tooth in comparison with the HAp phase PDF # 9-432. While this is the main identified phase in the diffraction patterns of all samples, minor unidentified phase(s) exist even in the young teeth, although within the detection limits of the method in the later. The identified secondary phases vary with the tooth-age qualitatively and quantitatively, as deduced from the Bragg peaks and their relative intensities. Poor crystallinity, broadening and overlapping of the diffraction peaks would make the phase identification ambiguous especially in old age teeth-samples. Possible secondary phases include the biologically relevant calcium compounds: $Ca_2(P_4O_{12}).4H_2O$ (#50-582), $Ca_8H_2(PO_4)_65H_2O$ (#26-1056), $Ca_6(CO_{2.65})_2(OH._{657}).7(H_2O)$ (#78-1540), $CaCO_3$ (#71-2392) and $Ca_2P_2O_7$ (73-440) [6,15]. Note that the type and number of the secondary phases vary in each specimen.

The x-ray diffraction patterns of Fig. 2 reveal a systematic decrease of the crystallinity of human dental apatite from 5 to 87 years old. It is quiet noticeable that patterns collected from teeth up to 45 years old show highly crystallized materials (with the exception of T43), whereas the patterns from older-age teeth display an increasing broadening of the Bragg peaks that indicates an increasing loss of crystallinity of the human dental apatite as a function of the tooth-age. The average crystallite size τ was calculated from the FWHM β of the (002), and (310) Bragg peaks using the Scherrer formula $\tau = \frac{K\lambda}{\beta \cos\theta}$. These two peaks were chosen because they do not overlap with others. It was found that the average crystallite-size in the specimens varies from ~12 nm (older age teeth) up to ~38 nm (younger age teeth). These numbers are in agreement with values found for crystallites in dentin [10] and enamel [16].

The average crystallographic properties of the specimens were found from Rietveld refinement of the powder diffraction patterns. One example is shown in Fig. 3 from the sample T38. The weighted $R$ factors of the refined patterns of all the samples were $0.15 \leq R_{wp} \leq 0.18$ except for the T87 with $R_{wp} = 0.28$. The reduced $\chi^2$ were $1.2 \leq \chi^2 \leq 1.6$ and the $R_{Bragg}$ were $0.06 \leq R_{Bragg} \leq 0.13$. The low counting rate, presence of secondary phases and nanoscale crystallite size explain such high residuals combined with low goodness of fit. The occupancies of the Ca and P sites refine to values less than one as expected from the chemical composition of dental apatite [1, 2].

A systematic decrease of the α-lattice constant with the tooth-age is demonstrated in Fig. 4. Decrease of the α lattice parameter in carbonate apatites is associated with an increase in carbonate content [1, 2, 17]. Higher number of planar carbonate ions substituting for the tetrahedral phosphate ions in the apatite structure is followed by an increased crystal structure disorder and reduction of the crystallinity as it is demonstrated by the broadening of the diffraction peaks in Fig. 2. This is biologically important because increase in carbonate content as a function of age also means increase of the solubility of the dental apatite [6] and consequently formation of calcium phosphate phases that alter the composition of the dental mineral. On the other hand, as Fig. 5 shows, no significant changes of the c-lattice constant with the tooth-age were found with the exception of one sample. Accordingly, no significant substitution variations occur at the channel (hydroxyl) site as a function of the tooth-age.

The refined interatomic distances between the atoms of the phosphate tetrahedron as calculated from the Rietveld refinements of the x-ray diffraction patterns are plotted in Fig. 6 versus the tooth-age. Notice that while in young age teeth up to ~40 years old the tetrahedral distances P-O2 (triangles) and P-O3 (circles) are distributed around the ideal value of 1.54 Å (marked with the dashed line), they show disturbance in older-age teeth. Moreover, the P-O1 distances (squares) are noticeably disturbed in all samples compared to the P-O2 and P-O3 bond lengths. Distortion of the phosphate tetrahedron is correlated with the well-known lattice disorder caused by the carbonate for phosphate (B-type) substitution in natural and synthetic apatites, referred as the "carbonate substitution problem" [18, 19]. Notice that this distortion is different from the one observed in carbonate natural fluorapatites and synthetic HAps that was studied earlier [17] in the sense that in those both the P-O1 and P-O2 interatomic distances of the atoms on the mirror plane of the phosphate tetrahedron were distorted by 3-4% because of the carbonate for phosphate substitution. Further investigation is required to draw conclusions on this subject regarding the dental carbonate HAp.

The wt % of the carbonate loss from several samples versus the tooth-age is plotted in Fig. 7, as evaluated from differential thermogravimetric analysis in the temperature range above 600 $^o$C up to 950 $^o$C. Weight losses of absorbed, adsorbed water or possible organic compounds that take place at temperatures less than 600 $^o$C were evaluated. An increase of carbonate content with the tooth-age is demonstrated in this figure confirming the correlation between the decrease of the α-lattice constant as a function of the tooth-age of Fig. 4 with increased carbonate content.

Fig. 8 shows the 800-1800 cm$^{-1}$ region of the IR spectra as collected from samples in the age range 5 to 86 years old. These spectra are characteristic of bio-apatites; the phosphate bands are identified by peaks at ~962 cm$^{-1}$ ($v_1$ PO$_4$ stretching IR mode), and the $v_3$ PO$_4$ region which appears as a very strong broad asymmetric band at ~1015 cm$^{-1}$ and consists of at least three sub-modes [20].

Strong peaks assigned to the B-type carbonate substitution (carbonate for phosphate ion) are observed at 872 cm$^{-1}$ ($v_2$ CO$_3$ mode) and at 1405, 1450 cm$^{-1}$ ($v_3$ CO$_3$). The weak bands in the $v_3$ CO$_3$ region are attributed either to CO$_3^{2-}$ replacing PO$_4^{3-}$ ions without an adjacent OH$^-$ ion [21] (at 1480 cm$^{-1}$), or to the A-type carbonate substitution [2] (weak shoulders at 880 cm$^{-1}$, ~1495 cm$^{-1}$ and ~1530 cm$^{-1}$).

Organic phase related bands, mainly due to dentin, have been observed in the Raman spectra of tooth samples [22, 23]. In particular, the Raman bands peaked at 1250, 1450 and 1670 cm$^{-1}$ were related with the amide III, amide II and amide I bands, respectively. The amide bands have been observed in the IR spectra of tooth samples above 1500 cm$^{-1}$ [24]. Therefore, the band at 1230 cm$^{-1}$ band is attributed to amide III. The broad feature above 1600 cm$^{-1}$ that consists of two sub-bands at 1610 and 1650 cm$^{-1}$ can be attributed to overlapping bands of carbonate containing phases other than HAp (carbonate probably at the channel sites) [2], with amide III bands. The 1650 cm$^{-1}$ band dominates over the 1600 cm$^{-1}$ in the spectra from teeth older than 45 years old. Usually the higher frequency sub-band is stronger at dentin untreated samples, while it loses intensity at enamel samples or upon treatment [24]. According to the above assignment the sub-bands behavior can be related either with the different content of dentin and enamel in the samples with age or with the secondary phases observed in the x-ray diffraction patterns. In favor of the secondary phases' explanation, some other weak bands at the $\nu_2$ $CO_3$ region also imply the presence of carbonate in slightly different environments than A- and B- type as mentioned above for the weak bands in the $\nu_3$ $CO_3$ region.

In a previous work we have used the ATR technique for a quantitative estimation of the relative carbonate content in specimens of synthetic and natural carbonate apatites [20] from the ratios of the intensities of the $\nu_2$ $CO_3$ modes to the $\nu_1$ $PO_4$. Fig. 9 (a) presents the ratio of the IR intensities of the $\nu_{2[B]}$ $CO_3$ mode to the $\nu_1$ $PO_4$ (B-type carbonate substitution) and Fig. 9 (b) the ratio of the $\nu_{2[A]}$ $CO_3$ to the $\nu_1$ $PO_4$ modes (A-type carbonate substitution). Fig. 9 clearly demonstrates a trend of increasing carbonate content with the tooth-age, which is in a good agreement with the results of the XRD experiments shown in Fig. 4 and the actual measurements of the carbonate loss of Fig. 7.

The maximum B- to A-type relative carbonate content is approximately 5, a value that is close to other estimates in biological apatites [9], [15]. Other authors [9] have found similar results by using the relative intensities of the 1415 cm$^{-1}$ ($\nu_{3[B]}$ $CO_3$) to the 603 cm$^{-1}$ ($\nu_4$ $PO_4$) band and the 1545 cm$^{-1}$ ($\nu_{3[A]}$ $CO_3$) to the $\nu_4$ $PO_4$ band respectively. We prefer to use the ratios as in Fig. 9 because there is no coexistence of more than one A- and B- type bands in the $\nu_2$ $CO_3$ region as in the $\nu_3$, hence we avoid a possible fitting procedure uncertainty.

**Conclusions**

Consistent, systematic variations of average crystal structure properties of human dental apatite as a function of age were found in this study from XRD, TGA and FTIR spectroscopy experiments. The decrease of the α-lattice constant versus age in dental apatite that is associated with increased carbonate content is related to increasing solubility which in turn results to a decrease of crystallinity and disturbance of the local lattice order of the biomineral. The approximately age-independent c-lattice parameter implies that the phosphate tetrahedron remains the main site of the carbonate substitution in the apatite lattice (B-type substitution) in the studied age-range. TGA measurements demonstrate an increased carbonate content with the tooth-age. FTIR spectra also show an increase of the B and A-type carbonate contents as a function of the age of the dental mineral phase with the B-type substitution up to 5 times greater than the A-type.

These trends of the average crystal structure properties of human dental apatite as a function of age could be useful in understanding the details of structural modifications in aging teeth. However, further research is required using specimens from a large, diverse pool in order to acquire statistical information considering that the tooth bioactivity is greatly affected by diet, diseases or other local factors that consequently affect the evolution of the mineral phase in aging human teeth.

**Acknowledgements**

We are grateful to Drs. J. R. Delacruz, C. A. Beck and J. R. Magnacca for providing the specimens for this study. We also thank Dr. L.A. O'Brien for kindly providing the deciduous tooth of her son. This work was supported by the Cancer Institute at the FAU Research Park, Boca Raton FL.

**Captions**

**Fig. 1.** X-ray diffraction pattern of the sample from the 17 year-old tooth (T17). All the peaks match the HAp pattern # 9-432 from the ICDS. A few minor peaks of unidentified phases are within detection limits of the method.

**Fig. 2.** Parts of the XRD patterns displaying the development of the HAp phase in human dental apatite as a function of the tooth-age.

**Fig. 3.** Rietveld refinement of the diffraction pattern collected from the 38 year old tooth (sample T38). Crosses mark the experimental data, the continuous line is the calculated HAp pattern, the vertical ticks mark the calculated Bragg peaks and the lower trace shows the difference between observed and calculated patterns.

**Fig. 4.** The $\alpha$-lattice constant of the HAp phase as a function of the tooth-age. Notice the increase of the error in poorly crystallized dental apatite (older-age teeth).

**Fig. 5.** The c-lattice constant as a function of the tooth-age.

**Fig. 6.** The refined interatomic distances between the atoms of the phosphate tetrahedron as a function of the human tooth-age. Squares mark the P-O1, triangles the P-O2 and circles the P-O3 bond lengths. Error bars are plotted for the P-O1 (not shown in this scale).

**Fig. 7.** The wt% of carbonate loss of human dental apatite as a function the tooth-age as evaluated from differential thermogravimetric analysis.

**Fig. 8.** The 800-1750 cm$^{-1}$ region of FTIR spectra of the human dental apatite measured from samples 5 to 86 years old.

**Fig. 9.** Ratios of the IR intensities of the $v_{2[B]}$ and $v_{2[A]}$ CO$_3$ modes to the $v_1$ PO$_4$ mode. **(a)** B-type and **(b)** A-type carbonate substitutions. An estimate of the relative carbonate content in human dental apatite as a function of the tooth-age is provided by these ratios.

Figure 1

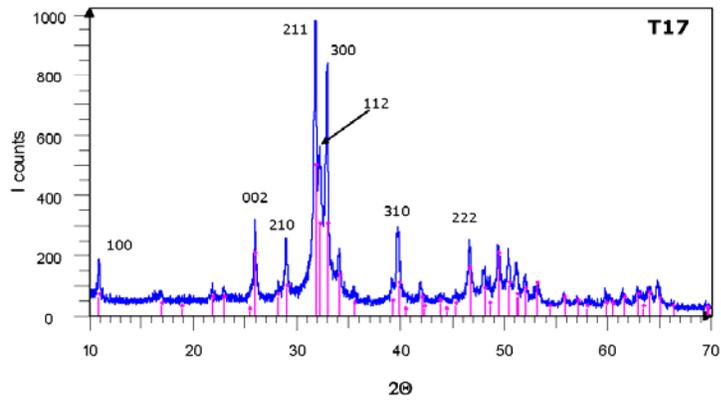

Figure 2

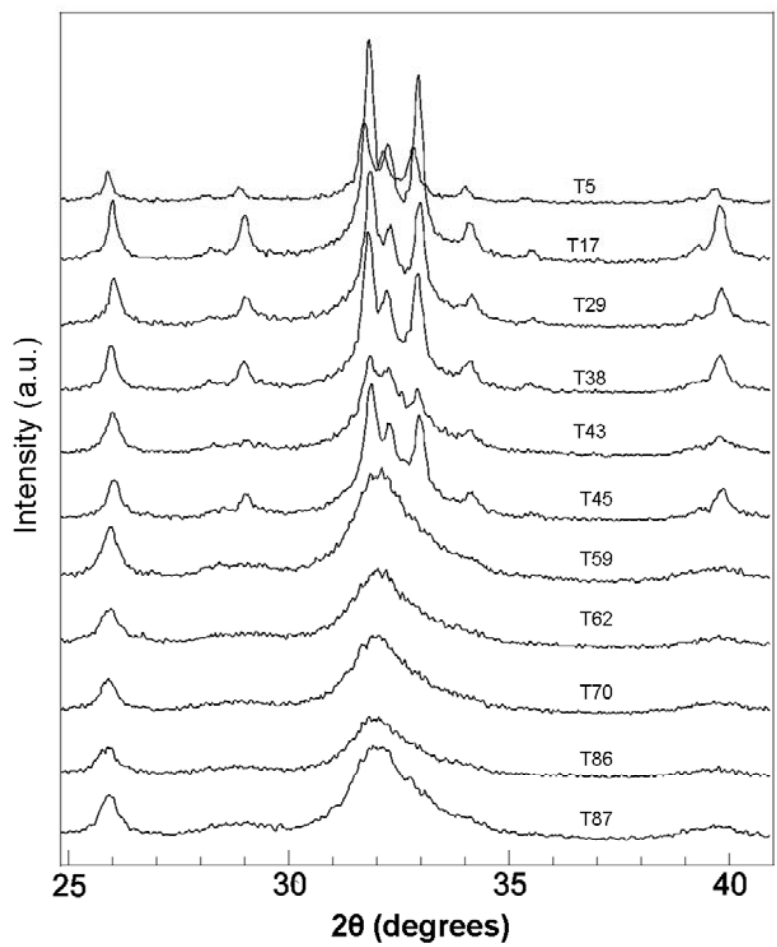

Figure 3

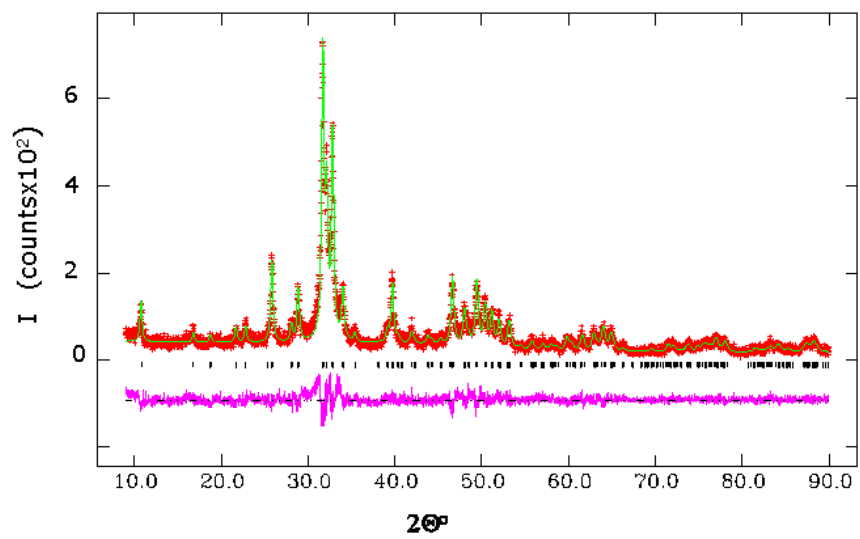

Figure 4

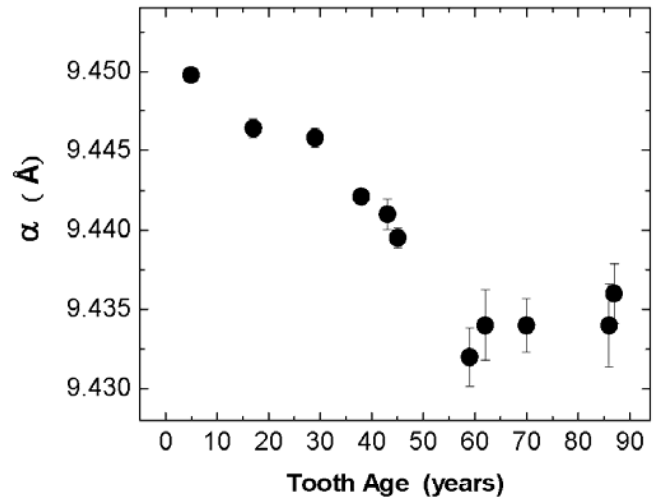

Figure 5

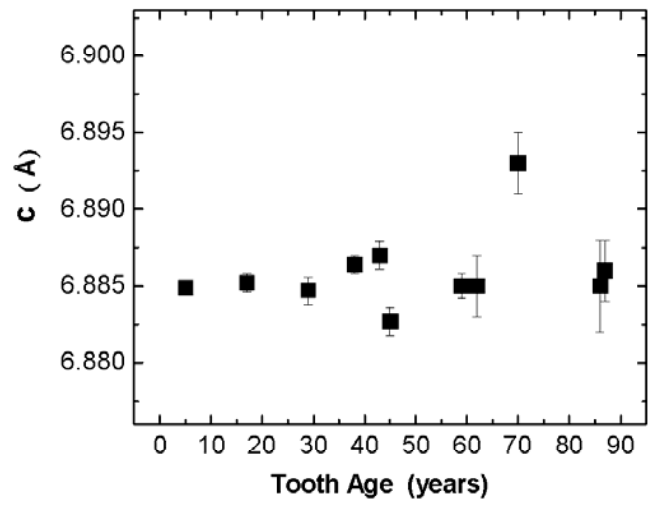

Figure 6

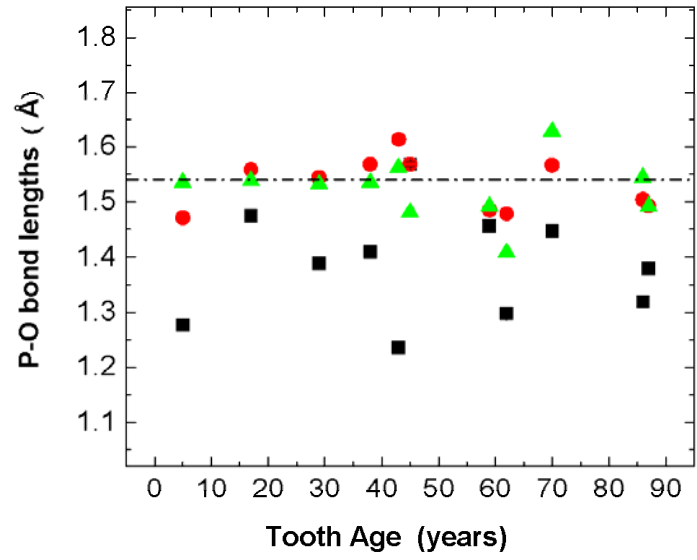

Figure 7

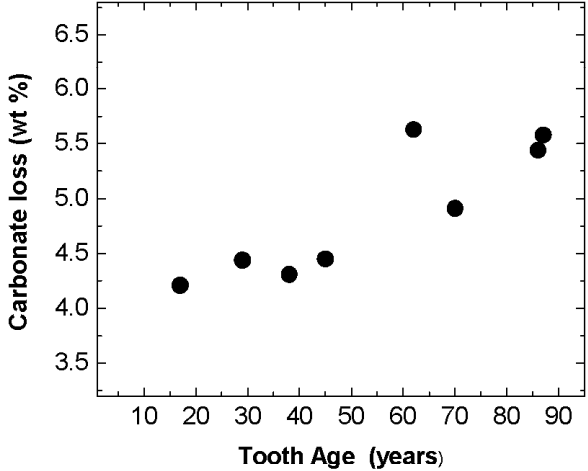

Figure 8

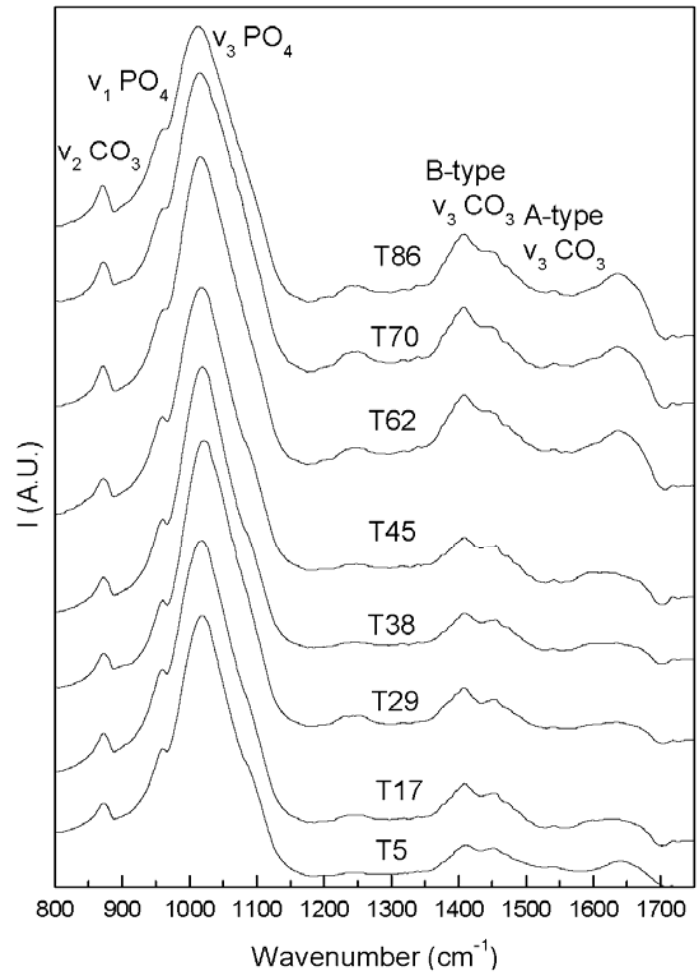

Figure 9

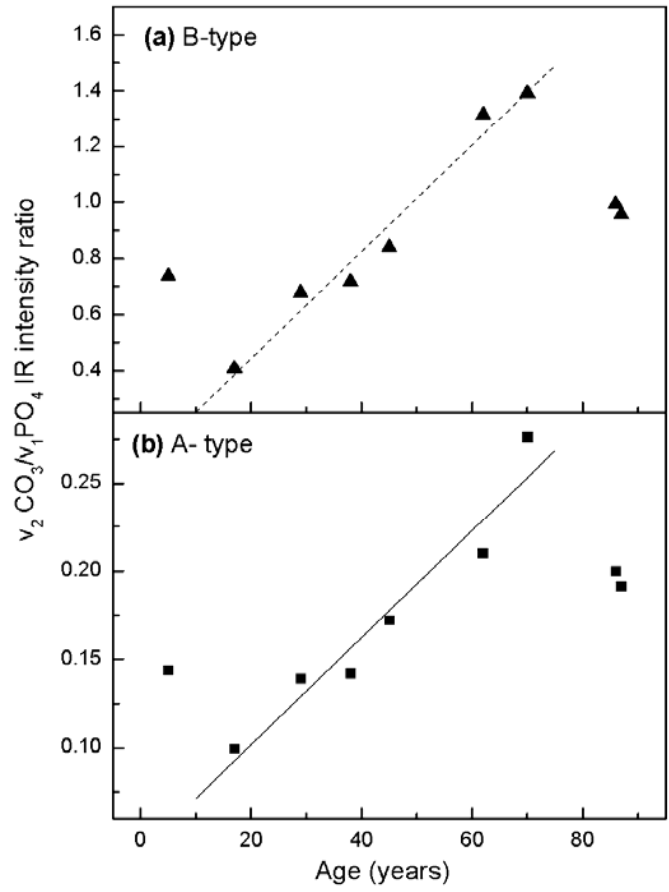

[19] Leventouri T. Synthetic and biological hydroxyapatites: crystal structure questions. Biomaterials 2006; 27:3339-42.
[20] Antonakos A, Liarokapis E, Leventouri T. Micro-Raman and FTIR studies of Synthetic and Natural Apatites. Biomaterials 2007; 28: 3043-3054.
[21] Vignoles-Montrejaud M. Contibution a 1'Etude des Apatites Carbonate es de Type B. These d'Etat, Institut National Polytechnique de Toulouse, 1984.
[22] Penel G, Leroy G, Rey G, Bres G. MicroRaman spectral study of the $PO_4$ and $CO_3$ vibrational modes in synthetic and biological apatites: Calcif. Tissue Int. 1998; 63:475–81.
[23] Tramini P, Pelissier B, Valcarel J, Bonnet B, Maury L. A Raman spectroscopic investigation of dentin and enamel structures modified by lactic acid, Caries Res. 2000; 34:233-240.
[24] Fattibene P, Carosi A, De Coste V, Sacchetti A, Nucara A, Postorino P, Dore P. A comparative EPR, infrared and Raman study of natural and deproteinated tooth enamel and dentin, Phys. Med. Biol. 2005; 50:1095–1108.

17